\documentstyle[12pt]{article}
\newcommand{\be}{\begin{equation}}
\newcommand{\ee}{\end{equation}}
\newcommand{\bea}{\begin{eqnarray}}
\newcommand{\eea}{\end{eqnarray}}
\newcommand{\al}{\alpha}

\newcommand{\gm}{\gamma}
\newcommand{\Gm}{\Gamma}
\newcommand{\dl}{\delta}
\newcommand{\Dl}{\Delta}
\newcommand{\eps}{\epsilon}

\newcommand{\lm}{\lambda}

\newcommand{\om}{\omega}
\newcommand{\pr}{\partial}
\newcommand{\pa}{\partial}
\newcommand{\dd}{\mbox{d}}

\newcommand{\uy}{\underline{y}}

\newcommand{\ux}{\underline{x}}

\newcommand{\uu}{\underline{u}}

\newcommand{\nn}{\nonumber}
\newcommand{\SH}{\hat{S}}
\newcommand{\Bbb}{\cal}

\begin{document}
\parindent=1.5pc
\begin{titlepage}

\begin{center}
{{\bf
Differential Renormalization, \\
the Action Principle and
Renormalization Group Calculations\footnote{Supported
by the Russian Foundation for Fundamental
Research, project 93-011-147.}}\\
\vglue 5pt
\vglue 1.0cm
{ {\large V.A. Smirnov}\footnote{E-mail: smirnov@compnet.msu.su } }\\
\baselineskip=14pt
\vspace{2mm}
{\it Nuclear Physics Institute of
Moscow State University}\\
{\it Moscow 119899, Russia  }\\
\vglue 0.8cm
{Abstract}}
\end{center}
\vglue 0.3cm
{\rightskip=3pc
 \leftskip=3pc
 \tenrm\baselineskip=12pt
\noindent
General prescriptions of differential renormalization are presented.
It is shown that renormalization group functions are
straightforwardly expressed through some constants that naturally
arise within this approach. The status of the action principle in the
framework of differential renormalization is discussed.
\vglue 0.8cm}
\end{titlepage}
\section{Introduction}

Differential renormalization
\cite{FJL,LMV} was invented as an alternative renormalization scheme
useful for calculations strictly in four dimensions \cite{appl,O}.
The basic idea of this renormalization\footnote{In \cite{P}
a renormalization
prescription of  differential style was much earlier formulated at the
level of primitively divergent diagrams, using the language of the
$\al$-representation.}
is to
represent products of propagators in coordinate
space\footnote{Euclidean and Minkowski spaces can be treated
on the same footing.
For simplicity in what follows Feynman amplitudes will be considered
in four-dimensional Euclidean space-time.}
\be
\Pi_{\Gm} (x_1, \ldots , x_N) =
\prod_{l} G(x_{\pi_+(l)} - x_{\pi_-(l)}),
\label{F0b}
\end{equation}
through derivatives of sufficient order acting on locally integrable
functions.
Here the product is over the lines of a given graph $\Gm$,
$\pi_{\pm}(l)$ are respectively beginning and the end of a line
$l$,
\be
 G(x) = P(\pa / \pa x, m) \:
\frac{m}{4\pi^2 \sqrt{x^2}} K_1 (m \sqrt{x^2})
\label{PROP}
\ee
is a propagator, with $P$ polynomial and $K_1$ a modified Bessel
function.
This procedure explicitly characterizes the $R$-operation (i.e.
renormalization at diagrammatic level)
as an extension of the functional $\Pi_{\Gm}(x_1, \ldots , x_n)$
from the subspace of test functions which vanish
in a vicinity of points where the
coordinates $x_i$ coincide to the whole space
${\cal D}({\cal R}^{4n})$.

The first step within initial version of differential
renormalization \cite{FJL,LMV} is to reduce the problem to the case of
diagrams depending on one coordinate difference.
To do this at low orders of perturbation theory, it suffices to exploit
certain
manipulations based on the Leibniz rule. At higher orders, the only way
of  performing such a reduction is to integrate over all coordinate
differences
except one.
However it is then possible to run into infrared problems
since this `naive' integration  generally induces infrared divergences.
In \cite{S}
the original version of differential renormalization was supplied
with simple prescriptions
which enabled infrared troubles to be avoided so that
differentially renormalized expressions could be found
with no more difficulty than determining the corresponding
counterterms
in dimensional renormalization. It was also shown that
in writing down
differentially renormalized quantities it is very useful to apply
calculational experience based on dimensional regularization.

The second step
\cite{FJL,LMV}
is performed with prescriptions of the following type:
\bea
\frac{1}{x^4}  &  \to &
- \frac{1}{4} \Box \frac{\ln \mu^2 x^2}{x^2},
\label{F1a}\\
\frac{\ln \mu^2 x^2}{x^4}  &  \to &
- \frac{1}{8} \Box
\frac{\ln^2 \mu^2 x^2 + 2 \ln \mu'^2 x^2}{x^2},
\label{F1b}\\
\frac{1}{x^6}   & \to &
- \frac{1}{32} \Box^2 \; \frac{\ln \mu^2 x^2}{x^2},
\label{F1c}
\eea
etc., where $\Box = \pa_{\al} \pa_{\al}$ is the usual Laplacian,
$x^4 = (x^2)^2, x^6 = (x^2)^3$, and
$\mu,\mu'$ are massive parameters which play the role of
subtraction points.
For $x \neq 0$, the expressions in the left-hand side and
the right-hand side of (\ref{F1a}--\ref{F1c}) are identical. By definition,
the extension of
functionals in the left-hand side from the subspace of test functions which
vanish near $x=0$ to the whole space is determined by the right-hand side.
Note that all the derivatives involved are understood in
the distributional sense, i.e. a derivative $D^{\al}f$ of
a distribution $f$ acts on a test function $\phi$ as
\be
(D^{\al}f,\phi) = (-1)^{| \al |}(f,D^{\al} \phi),
\label{DD} \ee
$| \al |$ being the order of the derivative.

In \cite{SZ} a second version of differential renormalization
was presented in the case of scalar massless logarithmically
divergent diagrams.
It was
based on `pulling out' another differential operator instead of the
Laplacian. In particular, (\ref{F1a}) is replaced by
\be
\frac{1}{x^4}   \to
\SH \: \frac{\ln \mu^2 x^2}{x^4},
\label{F1aa}
\ee
where
\be
\hat{S} = \frac{1}{2} \frac{\pa}{\pa x_{\al}} x_{\al}.
\label{SH0}
\ee
Within this version, there is no necessity of reducing the problem of
renormalization to propagator-type diagrams.
(This reduction is as usual important in
renormalization group calculations --- see below.) Thus there is no
asymmetry of treating  vertices of the given graph.

The purpose of this paper is to present a general prescription of
this version of differential
renormalization which is applicable for arbitrary diagrams
including massive ones.
The status
of the renormalized action principle within differential
renormalization will be also discussed. Another task is to show
that some constants that naturally arise within this approach \cite{SZ}
are straightforwardly related to the renormalization group coefficients.
It will be proved that the beta function and anomalous dimensions
are expressed through these constants
by the same formulae that,
in the case of the MS scheme, the RG coefficients are expressed
through counterterms.
Note that the differential renormalization happens to be a
mass-independent scheme.

The plan of the paper is as follows.
In the next section necessary differential operators similar to
(\ref{SH0})
will be presented and  standard formulae for the $R$-operation
will be listed. Then in Section 3 renormalization of massless lower
order diagrams is characterized. In Section 4 an auxiliary technique
necessary for renormalization in the massive case is introduced
through examples of
lower order graphs.
In Section 5 the general prescriptions are formulated and justified.
Section 6 is devoted to discussion of the action principle within
differential renormalization. In Section 7 explicit formulae for RG
coefficients will be derived. Finally, Section 8 contains discussion of
the results obtained.

\section{Notation}
\subsection{Differential operators}
Let us define the following differential operator:
\be
\SH_{\ux } = \frac{1}{2} \sum_{i=1}^{n}
\frac{\pa}{\pa x_{i \al}}
(x_{i \al} - \overline{x_{\al}}),
\label{F1}
\end{equation}
where $\ux  \equiv x_1, \ldots, x_n$ is a set of $n$
four-dimensional variables, and
$\overline{x} = \frac{1}{n} \sum x_i$.
If $F(\ux )$ is a translationally invariant function, i.e.
$F(\ux +a)=F(\ux )$, then
$F(\ux )=f(\uu )$ for
$u_i=x_i-x_{i_0}, \; i \neq i_0$ and
\be
\SH_{\ux }F(\ux )
=\hat{S}_{\uu }f(\uu ),
\label{F1ab}
\end{equation}
where
\be\hat{S}_{\uu } = \frac{1}{2} \sum_{i \neq i_0}
\frac{\pa}{\pa u_{i \al}} u_{i \al}.
\label{F2}
\end{equation}
Since the Feynman amplitudes are translationally invariant we will use
subsequently this form for the operator $\hat{S}$.

In fact, the homogeneity properties of Feynman amplitudes play an essential
role. If $f(u_1, \ldots, u_{n-1})$ is a homogeneous function or
distribution of degree $\lambda$ then
\[
\hat{S}f = \frac{1}{2} (\lambda + 4(n-1)) f .
\]

Note that the operator $\hat{S}$ involves a preliminary multiplication
by variables $u_i$ which vanish at points where initial
amplitudes are singular.
Correspondingly, these singularities are reduced.
In case the  ultraviolet
divergence is  logarithmic, it disappears if the subsequent
differentiation is understood in the distributional sense --- see
(\ref{DD}).

In the case of linear divergences multiplication by a monomial of the
first degree in coordinates is not sufficient. A second order monomial
is necessary so that it is natural to apply the following operator
\be\hat{S}^{(1)} = \frac{1}{4} \sum_{i,j,\al , \beta}
\frac{\partial}{\partial x_{i \al}} \frac{\partial}{\partial x_{j \beta}}
(x_{i \al} - \overline{x}_{\al}) (x_{j \beta} - \overline{x}_{\beta}) .
\label{F25}
\end{equation}

For massless graphs, it is sufficient to
apply (\ref{F2}), (\ref{F25}) and their generalizations. If
massive lines are present, we may use homogeneity of Feynman amplitudes with
respect to coordinates and inverse masses. Then a natural analog
of (\ref{F2}) is given by
\be\hat{S} = \frac{1}{2} \sum_{i} \frac{\pa}{\pa u_{i}} u_{i}
- \frac{1}{2} \sum_{l}
m_{l} \frac{\pa}{\pa m_{l}} ,
\label{SXM}
\end{equation}
since differentiation in masses also improves the ultraviolet
behaviour.

In the general case of degree of divergence $\omega$ let us apply
the following differential operator:
\be\hat{S}^{(\omega)} =
N \{ \underbrace{ \hat{S}^0 \ldots \hat{S}^0 }_{ \omega } \}  ,
\label{F29}
\end{equation}
where $\om$ is the degree of divergence,
$\hat{S}^0 \equiv \hat{S}$ is defined by (\ref{SXM}),
and the symbol of the $N$-product implies that all the derivatives
$\frac{\partial}{\partial u_{i}}$ are to the
left of all $u_{i'}$ while all the derivatives
$\frac{\partial}{\partial m_{j}}$ are to the right of $m_{j'}$.
It is not difficult to show that
\be\hat{S}^{(\omega)} = \hat{S} (\hat{S} + 1/2)
\ldots (\hat{S} + \omega/2 ) .
\label{F31}
\end{equation}
In particular
\bea
\hat{S}^{(1)} = \hat{S} (\hat{S} + 1/2),
\label{S1} \\
\hat{S}^{(2)} = \hat{S} (\hat{S} + 1/2 )
(\hat{S} + 1) .
\label{S2}
\eea

The following commutation relation will be also of use:
\be
\ln^{k} \mu^2 x^2 = \frac{1}{k+1} \left(
\hat{S} \ln^{k+1} \mu^2 x^2 -
\ln^{k+1} \mu^2 x^2 \hat{S} \right) .
\label{CR1}
\ee
Its generalization for $k=0$ in the case of the operator
$\SH^{(\om)}$ looks like
\be
1 = a_{\om} \SH^{(\om)} \ln \mu^2 x^2
- ( \ln \mu^2 x^2  - 4 b_{\om} ) (\SH + \om/2)
\label{CR1a}
\ee
and is understood in the sense that it acts on a quantity that
vanishes after the action of the square of the
operator $(\SH + \om/2)$ (in the
second term of the right-hand side the second and higher powers of this
operator are omitted). Here
\[ a_{\om} = (-2)^{\om}/\om!, \;
b_{\om} = 1 + 1/2 + \ldots + 1/\om. \]
Generalizations of (\ref{CR1a}) for arbitrary $k$ can be also derived
but we shall not write them explicitly.

Furthermore, we shall need the following commutation relation:
\be
\left(\SH_{\Gm} + \om_{\Gm}/2 \right) \Pi_{\Gm}
= \Pi_{\Gm \setminus \gm}
\left(\SH_{\gm} + \om_{\gm}/2 \right) \Pi_{\gm},
\label{CR2}
\ee
where $\Gm \setminus \gm$ denotes the subgraph which consists of lines
that do not belong to the subgraph $\gm$.

\subsection{$R$-operation}

Unrenormalized Feynman amplitudes are obtained from the
products $\Pi_{\Gm}$
by integrating over coordinates associated with internal
vertices:
\be
F_{\Gm}(x_1, \ldots, x_n) =
\int \dd x_{n+1} \ldots x_N \Pi_{\Gm} (x_1, \ldots, x_N) .
\label{FPI} \ee
The ultraviolet divergences manifest themselves through local
non-integrability of the function $\Pi_{\Gm}$. The $R$-operation
transforms this function into a locally integrable  function
$R\Pi_{\Gm}$ which therefore can be naturally regarded  as a
distribution.
The integration at large $x_i$ does not influence
the ultraviolet divergences so that when defining
$R\Pi_{\Gm}$ all the vertices can be treated as external.

As is well-known, the renormalization can be based either on subtractions
for each complete subgraph (a subgraph is called complete if,
in case it contains the endpoints of some line,
it necessarily contains the line itself, i.e. from $\pi_{\pm}(l) \in \gm$
it follows that $l \in \gm$),
or for all 1PI subgraphs.
The first type of renormalization was used in many early works on
renormalization --- see, e.g. \cite{BP,H} and is designed for
theories with Lagrangians and composite operators with normal
ordering.
The corresponding $R$-operation
acts on the Feynman amplitude
$\Pi_{\Gm}$ for the graph $\Gm$ as
\bea
R\Pi_{\Gm} = \sum_{{\cal V} =
{\cal V}_1 \cup \ldots \cup {\cal V}_j}
\Lambda({\cal V}_1) \ldots \Lambda({\cal V}_j) \Pi_{\Gm} \nonumber \\
\equiv R'\Pi_{\Gm} + \Lambda(\Gm) \Pi_{\Gm} .
\label{ROP}
\eea
The sum is over all decompositions of the set of vertices
$\cal V$ of the graph $\Gm$ into non-empty non-intersecting subsets
${\cal V}_1, \ldots, {\cal V}_j$. Moreover,
$\Lambda({\cal V}_i)$ is the counterterm operation for the subgraph
$\gm({\cal V}_i)$
composed of vertices ${\cal V}_i$ and all lines that are internal to
these vertices.
Remember that $\Lambda ({\cal V}_i)=1$ if $\gm({\cal V}_i)$ is an
isolated
vertex, and $\Lambda ({\cal V}_i)=0$, if $\gm({\cal V}_i)$
is not an 1PI divergent subgraph.

The operation $R'$
is called incomplete $R$-operation. This operation removes all
subdivergences of the diagram but does not include the overall
counterterm $\Lambda(\Gm)$.
This implies that the function
$R'\Pi_{\Gm}$ is locally integrable in the space of coordinates
except at the point where all the coordinates coincide.
Thus, the problem reduces to the extension of this function
to a distribution defined on the whole space.

For many reasons, a second type of renormalization is commonly
used.\footnote{For example, the scheme based on subtractions at zero
momenta is in the first case the BPH  renormalization \cite{BP,H} while
its analog of the  second type is the BPHZ renormalization \cite{Z}.
For dimensional renormalization, only the second type is
used in practice. In contrast to the first type,  it provides a
mass independent renormalization scheme.}
The corresponding $R$-operation looks like
\bea
R\Pi_{\Gm} = \sum_{\gm_1, \ldots, \gm_j}
\Delta(\gamma_1) \ldots \Delta(\gamma_j) \Pi_{\Gm} \nonumber \\
\equiv R'\Pi_{\Gm} + \Delta(\Gm) \Pi_{\Gm} .
\label{R1PI}
\eea
where $\Delta(\gm)$ is the corresponding counterterm operation, and
the sum is over all sets $\{\gm_1, \ldots, \gm_j\}$
of disjoint divergent 1PI subgraphs,
with $\Dl(\emptyset)=1$.

Note that these two types of renormalization coincide in the massless
case, due to zero values of massless vacuum diagrams.

\section{Lower order examples in the massless case}

In the case of graph of Fig.~1a
for $u \neq 0$ we have
\be\Pi_{1a}(u) \equiv
\frac{1}{16\pi^4} \frac{1}{u^4} =
\frac{1}{16\pi^4} \SH\, \frac{\ln \mu^2 u^2}{u^4} .
\label{F3}
\end{equation}
The left-hand side of (\ref{F3}) is ill-defined as a
distribution because this function is non-integrable in the vicinity
of the point $u = 0$. However the right-hand side is correctly defined
as a distribution everywhere in ${\Bbb R}^4$, since the operator $\hat{S}$
involves preliminary multiplication by $u_{\al}$, and the function
$u_{\al} / u^4$ is already locally integrable.
By definition, an extension of the functional in the
left-hand side to the whole space ${\Bbb R}^4$ is determined
with the help of the
right-hand side. The arbitrariness
of this extension is explicitly
contained in the parameter $\mu$ which plays the same role as the
corresponding parameters in the frameworks of dimensional and analytic
renormalizations.
Let us thus define the `differentially
renormalized' Feynman amplitude for the graph 1a by
\be
R \, \Pi_{1a} = \frac{1}{16\pi^4} \hat{S} \frac{\ln \mu^2 u^2}{u^4}
\label{F9}
\end{equation}
so that, in accordance with (\ref{DD}),
 the action of this distribution on a test function
$\phi(u) \in {\cal D} ({\Bbb R}^4)$ is defined by
\be
 (R \Pi_{1a}, \phi) = -
\frac{1}{16\pi^4}
\int {\rm d} u \frac{\ln \mu^2 u^2}{u^4}
u_{\al} \frac{\pa}{\pa u_{\al}} \phi(u).
\label{F9z}
\ee

The counterterm operation $\Dl (\Gm)$ for the graph 1a can be
`formally' represented as
\be
\Dl \Pi_{1a} = \frac{1}{16\pi^4} \left( \SH\, \frac{\ln \mu^2 u^2}{u^4}
- \frac{1}{u^4} \right) .
\label{F9a}
\ee
This quantity alone (as well as other counterterms and unrenormalized
or partially renormalized Feynman amplitudes) does not make sense as a
functional on the whole space of test functions.
However one can combine the sum of contributions of counterterm
operations into renormalized quantities in such a way that all the
obtained combinations will be sensible under integration.
It should be noted that the counterterm
(\ref{F9a}) vanishes for $u \neq 0$.

The functional $\hat{S} R \Pi_{1a}$
equals zero for $u \neq 0$ and therefore its support coincides with the
point $u = 0$. It is easy to verify that its action
on test functions that are zero at this point is zero.
Hence
\begin{equation}
\hat{S}R \Pi_{1a} = c_{1a} \delta^{(4)} (u) .
\label{F5}
\end{equation}
To calculate the constant $c_{1a}$ one may introduce analytic
regularization, to write (\ref{F9})  through
$\frac{\dd}{\dd \lm} \SH (x^2)^{\lm-2}$ at $\lm=0$ and apply the
expansion
\be
(x^2)^{\lm-2}  =  \frac{\pi^2}{\lm} \dl^{(4)} (x) + O(\lm^0),
\label{L}
\ee
with $\lm$ in the neighbourhood of the origin of the complex plane.
The result is
\be c_{1a} = \frac{1}{16 \pi^2}. \label{C1} \ee

The next example is the graph of Fig.~1b.
The subdivergence is removed according to prescription for the graph 1a.
The `incomplete' $R$-operation (i.e. without the last subtraction), when
applied to the Feynman amplitude under consideration, gives
\begin{equation}
R' \, \Pi_{1b} \equiv (1+\Dl(\gm_1)) \Pi_{1b}
= \frac{1}{(4 \pi^2)^4}
\frac{1}{v^2 (u-v)^2} \SH\, \frac{\ln \mu^2 u^2}{u^4} ,
\label{F6}
\end{equation}
where $\gm_1$ is graph 1a as a subgraph in 1b.

Using (\ref{CR1}) at $k=0$ one observes that if not all the coordinates
$u,v,0$ of the graph 1b coincide the following
equation is valid:
\begin{equation}
R' \, \Pi_{1b} = \frac{1}{(4 \pi^2)^4} \left\{
\SH_{u,v}\, \frac{\ln \mu'^2 v^2}{v^2 (u-v)^2} \SH_{u}\, \frac{\ln \mu^2
u^2}{u^4}
- \frac{\ln \mu'^2 v^2}{v^2 (u-v)^2} \,\hat{S}_u R \frac{\ln \mu^2
u^2}{u^4} \right\} .
\label{F7}
\end{equation}
Let us now use (\ref{F5}) and the equation
\begin{equation}
\frac{\ln \mu' v^2}{v^4} = \frac{1}{2}
\SH\, \frac{\ln^2 \mu'^2 v^2}{v^4}, \; \; v \neq 0 ,
\label{F7a}
\end{equation}
which enables us to differentially renormalize graph 1a with an additional
logarithm:
\begin{equation}
R \frac{\ln \mu' v^2}{v^4} \equiv R \ln \mu' v^2 \Pi_{\Gm/\gm_1}
=\frac{1}{2} \SH\, \frac{\ln^2 \mu'^2 v^2}{v^4}.
\label{F77a}
\end{equation}
After that, as for Fig.~1a, the right-hand side of (\ref{F7})
turns out to be defined as a functional on the whole space
${\cal D}({\Bbb R}^8)$. The differentially renormalized Feynman amplitude for
the graph 1b is defined as the corresponding extension of functional
(\ref{F7}) from the subspace of test functions vanishing in a
vicinity of the point $u=v=0$.

As a result we obtain
\bea
R \, \Pi_{1b} & = &
\hat{S} \ln \mu'^2 v^2 \, R' \Pi_{\Gm}
- c_{\gm_1} R \ln \mu'^2 v^2 \, \Pi_{\Gm / \gm_1}
\nn \\
& \equiv &
\frac{1}{(4 \pi^2)^4} \left\{
\SH_{u,v}\, \frac{\ln \mu'^2 v^2}{v^2 (u-v)^2} \SH_u \, \frac{\ln \mu^2
u^2}{u^4}
- \frac{1}{2} c_1 \SH_v \,
\frac{\ln^2 \mu'^2 v^2}{v^4} \delta (u) \right\},
\label{F8}
\eea
with $\Gm =$ 1b, $\gm =$ 1a.
The arbitrariness of the subtraction operation for the graph 1b itself
appears explicitly in the parameter $\mu'$.
It is possible, for example, to introduce a unique mass scale
$\mu$ that determines an energy scale for perturbation
theory and fix  all $\mu$-parameters which may arise as
$\mu_{\Gm} = \zeta_{\Gm} \mu$, with some constants $\zeta_{\Gm}$.
With this prescription $\mu$ determines a one-parametrical subgroup
of RG transformations
and is quite analogous, in its character, to
the 't~Hooft mass $\mu$ in dimensional renormalization.

The action of the counterterm operation for the graph 1b is formally
written as
\begin{equation}
\Dl \Pi_{1b} = R \Pi_{1b} - R' \Pi_{1b}  .
\label{F10}
\end{equation}
The counterterm (\ref{F10}) vanishes everywhere outside the point
$u = v = 0$.
Note that for all other points
\[
\hat{S} \, R \, \Pi_{1b} =
c_{1a} \hat{S} \Pi_{\Gm/\gm_1} .
\]
Therefore the left-hand and right-hand sides of this equation differ
by a functional with support localized at the point $u=v=0$.
Using the same arguments as in the first example we obtain
\bea
\hat{S} \, R \, \Pi_{1b}
& = &
c_{\Gm} \delta^{(8)}(u,v) +
c_1 \SH\, \frac{\ln \mu'^2 v^2}{v^4} \delta^{(4)} (u)
\nn \\
& \equiv &
c_{\Gm} \delta^{(8)}(u,v) + c_{\gm_1} R \Pi_{\Gm / \gm_1},
\label{F10a}
\eea
with some constant $c_{\Gm} \equiv c_{1b}$.

To calculate this constant let us integrate (\ref{F10a}) over $u$:
\be
c_{\Gm} \delta^{(4)}(v) = \frac{1}{(4 \pi^2)^4}
\left\{ \SH_{v} \frac{1}{v^2}
\int \dd u  \frac{1}{(u-v)^2} \SH_{u} \frac{\ln \mu^2 u^2}{u^4}
- \pi^2 \SH_{v} \frac{\ln \mu^2 v^2}{v^4} \right\} .
\label{2l1}
\ee
The integral in the braces can be evaluated by introducing analytic
regularization
\be
\int \dd u  \frac{1}{(u-v)^2} \SH_{u} \frac{\ln \mu^2 u^2}{u^4}
= \frac{\dd}{\dd \lm} \left. \left[ (\mu^2)^{\lm} \lm
\int \dd u  \frac{1}{(u-v)^2(u^2)^{2-\lm}}
\right] \right|_{\lm=0}
\label{2l2}
\ee
and applying one-loop massless formula in four dimensions
\be
\int \dd u  \frac{1}{(u^2)^{\lm_1}((u-v)^2)^{\lm_2}}
=\pi^2 G(\lm_1,\lm_2)
\frac{1}{(v^2)^{\lm_1+\lm_2-2}},
\label{2l3}
\ee
with four-dimensional $G$-function given by
\be
G(\lm_1,\lm_2) =
\frac{\Gm(\lm_1+\lm_2-2) \Gm(2-\lm_1) \Gm(2-
\lm_2)}{\Gm(\lm_1)\Gm(\lm_2)\Gm(4-\lm_1-\lm_2)} .
\label{2l4}
\ee
Therefore the first term in the braces in (\ref{2l1}) is rewritten as
\be
\pi^2 \SH_{v} \frac{1}{v^2}
\frac{\dd}{\dd \lm} \left. \left[ (1+\lm)
(\mu^2)^{\lm} (v^2)^{\lm-1} \right] \right|_{\lm=0} =
\pi^2 \SH_{v} \left[ \frac{1}{v^4} + \frac{\ln \mu^2 v^2}{v^4}
\right] .
\label{2l5}
\ee
As a result we obtain the value $c_{1b}= 1/(16\pi^2)^2$.

For Fig.~1c
using relation (\ref{F9a}) we have
\bea
R'\, \Pi_{1c} & = & [1 + \Dl(\gm_1)
+ \Dl(\gm_2) ] \Pi_{1c} \nn \\
  & = & \frac{1}{(4\pi^2)^4} \left\{
\frac{1}{(u-v)^4}  \SH\, \frac{\ln \mu^2 u^2}{u^4}
+ \frac{1}{u^4}  \SH\, \frac{\ln \mu^2 (u-v)^2}{(u-v)^4}
- \frac{1}{u^4 (u-v)^4} \right\},
\label{F11}
\eea
where $\gm_1$ and $\gm_2$ are respectively left and right simple
loops 1a in 1c.

As in the case of Fig.~1b, using equations (\ref{F5}),
(\ref{F7a}), (\ref{CR1})
and extending
functional (\ref{F11}) from the space of test functions determined in
${\Bbb R}^8$ with `deleted' point $u=v=0$ to the whole space
${\cal D}({\Bbb R}^8)$
we come to the following result:
\be
R \, \Pi_{1c}  =
\hat{S} \ln \mu'^2 v^2 \, R' \Pi_{\Gm}
- c_{\gm_1} (R \ln \mu'^2 v^2 \Pi_{\Gm / \gm_1}
+ R \ln \mu'^2 v^2 \Pi_{\Gm / \gm_2} ) .
\label{F12}
\ee

In the case of the `catseye' diagram Fig.~1d we have
\begin{equation}
R'\, \Pi_{1d}  =  [1 + \Dl(\gm_1) + \Dl(\gm_{21})
+ \Dl(\gm_{22}) ] \Pi_{1d} ,
\label{F13}
\end{equation}
where
\begin{equation}
(1 + \Dl(\gm_1)) \Pi_{1d} =
\frac{1}{(4\pi^2)^6} \frac{1}{u^2 v^2 (u-w)^2 (v-w)^2}
\SH\, \frac{\ln \mu_1^2 (u-v)^2}{(u-v)^4} ,
\label{F14}
\end{equation}
\bea
\Dl(\gm_{21}) \Pi_{1d} =
\frac{1}{(4\pi^2)^6} \frac{1}{(u-w)^2 (v-w)^2} \left\{
\SH\, \frac{\ln \mu_2^2 u^2}{u^2 v^2}
\SH\, \frac{\ln \mu_1^2 (u-v)^2}{(u-v)^4} \right. \nn \\
\left. - \frac{1}{2} c_1 \SH\,
\frac{\ln^2 \mu_2^2 u^2}{u^4} \delta (u-v)
- \frac{1}{u^2 v^2} \SH\, \frac{\ln \mu_1^2 (u-v)^2}{(u-v)^4}
\right\} ,
\label{F15}
\eea
and $\Dl(\gm_{22}) \Pi_{1d}$ is obtained by replacing
$u$ by $w-u$ and $v$ by $w-v$. Here $\gm_1$ is the central simple
loop; $\gm_{21}$ and $\gm_{22}$ are respectively left and right
graphs 1b as subgraphs of 1d.

Let us consider the space ${\Bbb R}^{12}$ with deleted origin $u=v=w=0$,
For each point in the vicinity of the origin we have at least one of the
following two possibilities:
({\em i}) $u \neq 0$ or/and $v \neq 0$,
({\em ii}) $u \neq w$ or/and $v \neq w$. We consider first case ({\em ii}).
Then the contribution from the counterterm of the subgraph $\gm_{22}$
disappears and $R'$ takes the form
\be
R'\Pi_{1d} =
\frac{1}{(4\pi^2)^6} \frac{1}{(u-w)^2 (v-w)^2}
R \Pi_{\gm_{22}} .
\label{R1d}
\ee
Using the
procedure described above, in particular (\ref{CR1}) at $k=0$,
(\ref{F5}) and (\ref{F10a}),
$R' \Pi_{1d}$ can be represented in the form
\be
R' \, \Pi_{\Gm} = \hat{S} \ln \mu^2 w^2 R' \, \Pi_{\Gm}
- c_{\gm_1} \ln \mu^2 w^2 R' \, \Pi_{\Gm / \gm_1}
- c_{\gm_{22}} \ln \mu^2 w^2 R' \Pi_{\Gm / \gm_{22}} .
\label{F15a}
\ee
The functional $\hat{S} \ln \mu^2 w^2 R \, \Pi_{\Gm}$ is naturally
extended to the whole ${\Bbb R}^{12}$; the extension of simple loops
(with `additional' logarithms)
$\ln \mu^2 w^2 R' \, \Pi_{\Gm / \gm_{2i}}, \; i=1,2$
was described in the case of Fig.~1b. For the graph of Fig.~1c with
an additional
logarithm $\ln \mu^2 w^2 R' \, \Pi_{\Gm / \gm_1}$,
the procedure that was used for
Fig.~1c itself and applies (\ref{CR1}) at $k=2$ can be straightforwardly
generalized. A similar expression is obtained for the case ({\em i}).

As a result we obtain the following expression for the differentially
renormalized diagram 1d which is valid in the whole space:
\be
R \, \Pi_{\Gm} = \hat{S} \ln \mu^2 w^2 R' \, \Pi_{\Gm}
- c_{\gm_{1a}} R \ln \mu^2 w^2 \Pi_{\Gm / \gm_1}
- c_{\gm_{1b}} R \ln \mu^2 w^2 \Pi_{\Gm / \gm_{21}}
- c_{\gm_{1b}} R \ln \mu^2 w^2 \Pi_{\Gm / \gm_{22}}.
\label{F16}
\ee

\section{Lower order examples in the massive case}

Renormalization of the diagrams of Fig.~1 in the massive case is performed
by formulae similar to the previous section, with
$\SH$ given by (\ref{SXM}), e.g. for
graph 1a
\begin{equation}
R \, \Pi_{1a} \equiv R G(x)^2
= \hat{S} \ln \mu^2 x^2 G(x)^2.
\label{F99}
\end{equation}
The constants $c_{\gm}$ involved here
have the same values as in the massless case.

To see the additional problems that can arise
let us consider other simple examples.
New problems
can appear for vacuum graphs or those with one external vertex,
e.g. for tadpoles.
Let us distinguish contributions (see Fig.~2a) which are formally given
by $G(0)$
--- the value of propagator at the origin where it is singular.
They can
exist separately or belong to other graphs. In the latter
case, such contributions appear as independent factors. All the other tadpoles
are products of propagators integrated over all coordinates but one.
The methods of differential renormalization can be  applied for these
tadpoles without problem.

Note that even within dimensional regularization
$G(0)$ is not defined without some further prescription.
A standard way of handling this quantity
is write it down as momentum space
$d$-dimensional integral
$\int \mbox{d}^d k/(k^2+m^2)$, then calculate
and renormalize it. In the limit $d\to 4$,
one obtains $G(0)=\frac{1}{16\pi^2}m^2\ln m^2/\mu^2$
(up to finite renormalization).

At $d=4$ it is possible to say that $G(0)$ is understood as the value at
$x=0$ of $G(x)$
from which the singular terms of the small $x$ expansion
\be
G(x) = \frac{1}{4\pi^2} \left\{ \frac{1}{x^2}
+ \frac{m^2}{4}\left( \ln m^2 x^2 + 2 \gm_{\rm E} -1
 \right) \right\} + o(x^2),
\ee
are subtracted
($\gm_{\rm E}$ the Euler constant). This procedure
leads to the reasonable result
$G(0)=\frac{1}{16\pi^2}m^2 \left( \ln m^2/\mu^2 + 2 \gm_{\rm E} -1\right)$.

At first sight the strategy of extending functionals from some space of test
functions
to the whole space does not have anything to do with defining $G(0)$.
Therefore the general recipes of differential renormalization seem to be
of no use here. One possibility is to introduce
additional prescriptions, and there can be  at least
 two variants: to set $G(0)=0$ or to take
$G(0)=\frac{1}{16\pi^2}m^2\ln m^2/\mu^2$ (or some similar value for non-scalar
propagators) --- the former variant of course reduces to the latter one at
$\mu=m$.

There is also a new problem of another type in the massive case. Consider
the `setting sun' diagram shown in Fig.~2b.
In the case of
renormalization (\ref{ROP}) based on complete subgraphs
the only subtraction involved is for the overall graph.
However the operator (\ref{S2})
is not sufficient to remove all the ultraviolet divergences because
differentiation in the mass removes only the most singular part of
$G(x)^3$.
If one tries to perform the second type of renormalization
(\ref{R1PI}) in the
differential style
one will observe that the insertion of counterterms for three (overlapping)
simple loops does not change the diagram at $x \neq 0$. Furthermore,
in the whole space, insertion of these counterterms does not make
sense. It seems that we do not have `enough space' to perform consecutive
extensions of the initial functional in two steps that correspond
respectively to renormalization of above three subgraphs and the graph
itself.

To overcome these problems let us exploit the following trick that was
used in \cite{L}. Instead of propagators in coordinate space
(\ref{PROP}), let us consider its Fourier transform in respect to two
additional components, $m_1$ and $m_2$ by considering the square of
mass $m^2$ in the propagator as the square of this two-dimensional
vector $m=(m_1,m_2)$.
Denoting the corresponding coordinate-space variable by
$y=(y_1,y_2)$ we obtain the massless propagator in six-dimensional
space
\be
G(x,y) = \frac{1}{4\pi^3} \frac{1}{(x^2+y^2)^2}
\label{Lu}
\ee
which satisfies
\be
(\Box_x + \Box_y) G(x,y) = - \dl^{(4)}(x)\dl^{(2)}(y).
\ee
For general Feynman diagram let us introduce Fourier transformation
for each massive line. As a result $\Pi_{\Gm}$ is expressed
as a product of propagators $G(x,y)$ depending on $4N$ usual
coordinate-space variables and $2N_m$ additional ones, $N_m$ being the
number of massive lines.

Remember that
Feynman amplitudes must be well-defined distributions in order that
Fourier transforms to momentum space
(where physical quantities are calculated)
are possible.
Therefore it is necessary
to perform integration over coordinates; at this step locally
non-integrable singularities manifest themselves as the source of the
ultraviolet divergences. Now, for the same reason, it is natural to
consider the
product of propagators $\Pi_{\Gm}(\ux, \uy)$, with
$\uy = (y_1, \ldots, y_{N_m})$,
as a distribution not only in $x_i$ but also in $y_i$. Indeed, in
the end it is necessary to perform inverse Fourier transformation and
come back to masses (and put them equal to each other if there was
initially only one mass).
In this approach the ultraviolet divergences manifest
themselves in integrations over small coordinate differences {\em and}
also the variables $y_i$.

Note that it is possible to introduce Fourier
transformation in masses even for massless lines. In this case, it is
necessary, in the end of calculation, to integrate over the
corresponding $y$-variables.
Therefore, as products of propagators, we now have
\be
\Pi_{\Gm} (\ux,\uy) =
\prod_{l} \frac{1}{4\pi^3}
\frac{1}{((x_{\pi_+(l)} - x_{\pi_-(l)})^2+y_l^2)^2} .
\label{PXY}
\end{equation}

One can show that usual power counting turns out to be
the same and is governed by the same degree of divergence $\om$.
It is now possible to apply above formulae of differential
renormalization (for the massless case) using the operator
\begin{equation}
\hat{S}_{x,y} = \frac{1}{2}
\left( \sum_{i} \frac{\pa}{\pa x_{i }} x_{i }
+ \sum_{l} \frac{\pa}{\pa y_{l}} y_{l} \right) .
\label{SXY}
\end{equation}

We can now write the value of the propagator at $x=0$
\be
G(x,y)\big |_{x=0} = \frac{1}{4\pi^3} \frac{1}{(y^2)^2} ,
\label{G0}
\ee
without running into division by zero. To make sense of it for
all $y$,
i.e. as distribution in $y$,
let us apply the operator (since the divergence is quadratic)
$\SH^{(2)}_{y}$ given by (\ref{S2}):
\be
R G(0,y) = \frac{1}{4\pi^3} 2\SH^{(2)}_y \frac{\ln \mu^2 y^2}{(y^2)^2} .
\label{G00}
\ee
Its inverse Fourier transformation in $y$ gives
\be
R G(0;m) = \frac{1}{16\pi^2}m^2
\left( \ln m^2/4\mu^2 + 1 + 2\gm_{\rm E} \right) .
\label{G0m}
\ee

The renormalized simple loop (\ref{F99}) is rewritten in the
language of $y$-variables as
\begin{equation}
R \, \Pi_{1a} = \frac{1}{(4\pi^3)^2} \hat{S}_{x,y_1,y_2}
 \frac{\ln \mu^2 V}{(x^2+y_1^2)^2 (x^2+y_2^2)^2} ,
\label{F90}
\end{equation}
with the corresponding counterterm represented as
\be
\Dl \Pi_{1a} = \frac{1}{(4\pi^3)^2}
\left[ \hat{S}_{x,y_1,y_2}
\frac{\ln \mu^2 V}{(x^2+y_1^2)^2 (x^2+y_2^2)^2}
- \frac{1}{(x^2+y_1^2)^2 (x^2+y_2^2)^2} \right] .
\label{F90a}
\ee
One can use different arguments of the logarithm involved, for
instance, (a) $V=x^2$; (b) $V = y_1^2$ or $V=y_2^2$,
(c) $V = x^2+y_1^2$ or $V = x^2+y_2^2$; (d) $V = x^2+y_1^2+y_2^2$.
It is possible to show that in cases (a) and (b) the limit $m\to 0$ exactly
reproduces the `pure massless' prescription (\ref{F9}).

Let us now return to the `setting-sun' diagram.
In the language of $y$-variables the product of
corresponding propagators is
\be
\Pi_{2b} = \prod_{i=1,2,3} \left(
\frac{1}{4\pi^3} \frac{1}{(x^2+y_i^2)^2} \right) .
\label{2a}
\ee
Inserting counterterms for three  subgraphs 1a gives
\be
R'\Pi_{2b} = \Pi_{2b}
+ G(0,y_3)
\left[ \hat{S}_{x,y_1,y_2}
\ln \mu^2 x^2 G(x,y_1) G(x,y_2) - G(x,y_1) G(x,y_2)
\right] + \ldots ,
\label{R'2a}
\ee
where the dots stand for permutations.
This quantity is  meaningful as a functional everywhere except the
origin with respect to the variables $(x,y_1,y_2,y_3)$.
Let us now apply (\ref{CR1a}) at $\om=2$ and use
the values of one-loop counterterms to write down the incompletely
renormalized diagram as
\be
R'\Pi_{2b} = 2\SH^{(2)} \ln \mu^2 V \; R' \Pi_{2b}
- \frac{1}{4\pi^3} (\ln \mu^2 y_3^2 -6) \frac{1}{(y_3^2)^2}
\frac{1}{16\pi^2} \dl^{(4)}(x) \dl^{(2)}(y_1) \dl^{(2)}(y_2) -\ldots ,
\label{R'2a1}
\ee
with $V$ chosen as $V = x^2 + y_1^2 +y_2^2 +y_3^2$.
Now we extend this functional to the whole space of variables
$(x,y_1,y_2,y_3)$ with the help of the renormalization of the $G(0)$
(see (\ref{G00})) and its analog with an additional logarithm:
\be
R \frac{\ln \mu^2 y^2}{y^4} =
\SH^{(2)} \; \frac{\ln^2 \mu^2 y^2 + 6 \ln \mu^2 y^2}{y^4}.
\label{PL}
\ee
We obtain
\bea
R\Pi_{2b} & = &
2\SH^{(2)} \ln \mu^2 V \; R' \Pi_{2b} \nn \\
&&{} - \frac{1}{4\pi^3} \SH^{(2)}_{y_1}
\frac{\ln^2 \mu^2 y_1^2 -6\ln \mu^2 y_1^2}{y_1^4}
\frac{1}{16\pi^2} \dl^{(4)}(x) \dl^{(2)}(y_2) \dl^{(2)}(y_3) - \ldots ,
\label{R2a}
\eea
where we did not use a freedom to introduce a new parameter $\mu$
associated with the overall graph.

Let us now calculate the constants $c_{\gm}$ which enter the following
formula:
\bea
(\SH+1) R\Pi_{2b}  & = &
(c_x \Box_x + \sum_i c_{y_i} \Box_{y_i}) \dl^{(4)}(x)
\prod_{i} \dl^{(2)}(y_i) \nn \\
&& {}+ R \ln \mu^2 y_1^2
G(0,y_3) c_{1a} \dl^{(4)}(x) \dl^{(2)}(y_2) \dl^{(2)}(y_3)
+ \ldots .
\label{S+1}
\eea

To calculate $c_x$ we multiply (\ref{S+1}) by $x^2$ and integrate it
over $x$ and $y_i$.
To calculate $c_y$ we integrate (\ref{S+1})
over $x,y_2,y_3$ and perform inverse Fourier transformation in $y_1$.
The results are
\be
c_x = \frac{1}{2} \frac{1}{(16 \pi^2)^2} , \;
 c_{y_i} = \frac{2}{(16 \pi^2)^2}.
\label{cuy}
\ee

\section{General prescriptions}

By generalizing procedure described in the above examples let us
define a renormalization procedure which will be naturally called as
differential renormalization.
To present quite general prescription let us apply the language of
$y$-variables described in the previous section even in the case when
some of the lines are massless. Let us consider renormalization
of products (\ref{PXY}) multiplied by powers of logarithms:
$ \ln^k (\mu^2_{\Gm} V_{\Gm}(\ux,\uy)) \, \Pi_{\Gm}$.
Here we imply two possible variants:
$V_{\Gm}= u^2_{\Gm} + \sum_{l \in \Gm} y_l^2$
and
$V_{\Gm}= \sum_{l \in \Gm} y_l^2$, where
$u_{\Gm}=x_i-x_i'$ is any difference variable of the considered
Feynman amplitude.

{\bf Definition.}
Let a renormalization $R$ of the graph $\Gm$ be given by the following
recursive formulae:
\be
R\,  \ln^k (\mu^2_{\Gm} V_{\Gm}) \Pi_{\Gm}  =
\frac{1}{k+1}
\hat{S}  \ln^{k+1} (\mu^2_{\Gm} V_{\Gm})
\; R'\, \Pi_{\Gm} -
\frac{1}{k+1} \sum_{\gm \subset \Gm}
R \ln^{k+1} (\mu^2_{\Gm} V_{\Gm}) \, {\cal C}_{\gm}
\Pi_{\Gm},
\label{F17}
\ee
for $\om=0$ and integer $k \geq 0$;
\be
R\,\Pi_{\Gm}  =
a_{\om} \SH^{(\om)} \ln (\mu^2_{\Gm} V_{\Gm})
\; R'\, \Pi_{\Gm} -
\frac{1}{k+1} \sum_{\gm \subset \Gm}
R ( \ln (\mu^2 V_{\Gm})   - 4 b_{\om} ) (\SH + \om/2)
{\cal C}_{\gm}
\Pi_{\Gm},
\label{F17a}
\ee
for $\om>0$ and integer $k = 0$, as well by other relations for arbitrary
$\om$ and $k$ which follow from the corresponding generalizations of
(\ref{CR1a}).
Here $R'$ is incomplete $R$-operation (\ref{ROP}).
The sum in the second term of the right-hand side of (\ref{F17}) and
(\ref{F17a}) runs over all 1PI
proper subgraphs $\gm$ of $\Gm$.
Furthermore,
\be
\Dl(\Gm) \Pi_{\Gm} = R \Pi_{\Gm} - R'\, \Pi_{\Gm}
\label{Dl}.
\ee
Finally, the operations ${\cal C}_{\gm}$
are determined from equations
\bea
\left( \hat{S} + \om /2 \right) R\, \Pi_{\Gm} =
\sum_{\gm \subseteq \Gm}
{\cal C}_{\gm} R \Pi_{\Gm}
\equiv \sum_{\gm \subset \Gm}
R {\cal C}_{\gm} \Pi_{\Gm }
+ {\cal C}_{\Gm} \Pi_{\Gm },
\label{F18}
\eea
where the operation
${\cal C}_{\gm}$ inserts a polynomial ${\cal P}_{\gm}$
of degree $\om(\gm)$ in masses
of $\gm$ and its external momenta into the reduced vertex of the graph
$\Gm /\gm$. Symbolically we write
\be
{\cal C}_{\gm} \Pi_{\Gm } = \Pi_{\Gm /\gm} \circ {\cal P}_{\gm}
\label{CG}
\ee
where $\circ$ denotes the insertion operation. In the language of
coordinate space,
\be
{\cal C}_{\Gm} \Pi_{\Gm } (\ux,\uy) =
{\cal P}_{\Gm} (\pa /\pa x_i, \pa /\pa y_l) \prod_{i \in \Gm}
\dl^{(4)}(x_i-x_1) \prod_{l \in \Gm} \dl^{(2)}(y_l) .
\label{CGC}
\ee
Indeed relations (\ref{F17})--(\ref{F18}) and their
generalizations for arbitrary $\om$ and $k$ enable us to obtain
$R \ln^k (\mu^2_{\Gm} V_{\Gm})\;
\Pi_{\Gm}$, $\Dl(\Gm)$ and ${\cal C}_{\Gm}$ provided we know the corresponding
quantities for all proper subgraphs of $\Gm$ and its reduced graphs:
in particular, ${\cal P}_{\Gm}$ is expressed from (\ref{F18}).

The following proposition is valid.

{\bf Proposition.} The procedure $R$ defined by relations
(\ref{F17})--(\ref{F18}) is a correct $R$-operation.

{\em Proof.}
To prove this proposition we should show that ({\em i}) the expression
(\ref{F17}) is finite, ({\em ii}) (\ref{F17}) is  obtained from
$R'\Pi_{\Gm}$ as an extension from the space from deleted origin
of the whole space, i.e. the corresponding counterterm
(\ref{Dl}) is local,
({\em iii}) the quantity $ {\cal C}_{\Gm} \Pi_{\Gm }$ found from
(\ref{F18}) is local.

Let us consider, in the space of coordinates, an arbitrary point
$\underline{x^0} \equiv \{ x^0_1, \ldots, x^0_n\}$ in which at least
one of the difference variables (e.g. $u^0_1 = x^0_1-x^0_n$) is
non-zero: $u^0_1 \neq 0$.
In respect to the point
$\underline{x^0}$ all the set of vertices $\cal V$ of the graph
$\Gm$ is naturally decomposed over non-intersecting subsets
${\cal V}_{r}$, with ${\cal V} = \bigcup_{r} {\cal V}_{r}$,
$x^0_i - x^0_{i'} =0, \; \forall i,i' \in {\cal V}_{r}$, and
$x^0_i - x^0_{i'} \neq 0, \; \forall i \in {\cal V}_{r}, \;
i' \in {\cal V}_{r '}$ for $r \neq r'$. In accordance with the
assumption about the point $\underline{x^0}$, the number of subsets
${\cal V}_{r}$ is not less than two.

Let $\Gm_{r}$ be subgraphs constructed with help of vertex sets
${\cal V}_{r}$: by definition, each $\Gm_{r}$ contains any line
that connects a pair of vertices from ${\cal V}_{r}$. Note that the
subgraphs $\Gm_{r}$ can be one-particle-reducible or disconnected.
Let us denote by $\cal W$ the set of maximal 1PI subgraphs of the
graph $\Gm^0 = \bigcup_{r} \Gm _{r}$. Furthermore, let
$\overline{\cal W}$ be the set of all divergent subgraphs of the
graph $\Gm^0$.

Let $U^0$ be a sufficiently small
vicinity of this point so that for all $\ux  \in U^0$
the same properties hold,
$x^0_i - x^0_{i'} =0, \; \forall i,i' \in {\cal V}_{r}$, and
$x^0_i - x^0_{i'} \neq 0, \; \forall i \in {\cal V}_{r}, \;
i' \in {\cal V}_{r '}$ for $r \neq r'$.
In the domain $U^0$, the incomplete $R$-operation $R'$ does not include
counterterms contributed by subgraphs containing vertices from
different subsets ${\cal V}_{r}$. Therefore, in $U^0$, we have the
equation
\begin{equation}
\hat{S} R' \, \Pi_{\Gm} =
\hat{S} \{ \sum
\Dl(\gm_1) \ldots \Dl(\gm_j) \}
\Pi_{\Gm},
\label{FA2}
\end{equation}
where the sum is over decompositions
${\cal V} = {\cal V}_1 \cup \ldots \cup {\cal V}_j$ such that any
$\gm_i$ belongs to $\overline{\cal W}$. In other words, the sum in
(\ref{FA2}) can be represented as analogous sum over decompositions in
which any $\gm_i$ happens to be an element from the set $\cal W$.
After that each of the factors $\Dl(\gm_i)$ involved transforms
into the $R$-operation $R(\gm_i)$ that acts on the Feynman
amplitude for the subgraph $\gm_i$. Thus,
(commutation relations (\ref{CR2}) are used)
\be
(\hat{S}+\om_{\Gm}/2) R' \, \Pi_{\Gm} =
\hat{S}
\left\{ \prod_{\gm \in {\cal W}} (R \Pi_{\gm}) \right\}
\Pi_{\Gm \setminus \Gm^0}
= \sum_{\gm \in {\cal W}}
((\hat{S}+\om_{\gm}/2) R \Pi_{\gm})
) \prod_{\gm' \neq \gm} ( R \Pi_{\gm '}) \; \Pi_{\Gm \setminus \Gm^0}.
\label{FA3}
\ee

Since in $U^0$ the propagators of lines which connect different
elements of ${\cal W}$ are not singular we may apply the Leibniz
rule encoded in (\ref{CR1}), (\ref{CR1a}) and their generalizations. For
example, in the  case $\om=0$ we have
\be
\ln^k (\mu^2_{\Gm} V_{\Gm}) \; R' \Pi_{\Gm}  =
\frac{1}{k+1}
\hat{S}  \ln^{k+1} (\mu^2_{\Gm} V_{\Gm})
\; R'\, \Pi_{\Gm} -
\frac{1}{k+1}
\ln^{k+1} (\mu^2_{\Gm} V_{\Gm})\, \SH R' \Pi_{\Gm}.
\label{LR}
\ee

Let us now apply relation (\ref{F18}) for subgraphs and turn
from summation over elements $\gm \in {\cal W}$ and subgraphs of
each $\gm$ to summation of subgraphs of $\Gm^0$. After that we
obtain the following expression for the second term in the
right-hand side of (\ref{LR}):
\be
\frac{1}{k+1}
\ln^{k+1} (\mu^2_{\Gm} V_{\Gm}) \; \sum_{\gm \subseteq {\cal W}}
R' {\cal C}_{\gm} \Pi_{\Gm}.
\label{LR1}
\ee
Here only subgraphs and reduced graphs with a smaller number of
loops are involved. Therefore we know how to renormalize these
quantities
$\ln^{k+1} (\mu^2_{\Gm} V_{\Gm}) \, R' {\cal C}_{\gm} \Pi_{\Gm}$
by extending them as functionals to the whole space and arriving
at
$R \ln^{k+1} (\mu^2_{\Gm} V_{\Gm}) \; {\cal C}_{\gm} \Pi_{\Gm}$.
As to the first term in (\ref{LR}) it does not have divergences
because all subdivergences are removed by $R'$ and the overall
divergence is removed by the operator
$\SH^{(\om)}$.
Thus we arrive at a finite differentially renormalized quantity
(\ref{F17}) which is obtained by extension of the functional
$R'\Pi_{\Gm}$ to the whole space (i.e. by adding a local
counterterm).
Note that in (\ref{F17})  the summation is all divergent 1PI
subgraphs of $\Gm$; in each vicinity $\Gm^0$ this summation
reduces to the corresponding set ${\cal W}$.

To prove ({\em iii}) it is sufficient to repeat the same
manipulations as for ({\em ii}) starting from
$(\hat{S}+\om/2)  R'\, \Pi_{\Gm}$
instead of
$\hat{S} \ln^{k+1} (\mu^2_{\Gm} V_{\Gm})\; R'\, \Pi_{\Gm}$.

{\em Comments.} ({\em a}) For renormalizable theories in
the pure massless case
there is no need to Fourier transform to $y$-variables. One can choose
$V_{\Gm}=u^2_{\Gm}$ where
$u_{\Gm}=x_i-x_i'$ is any difference variable of the considered
Feynman amplitude
 such that the vertices $i$  and $j$ do not belong to
the same divergent subgraph --- see examples in Section 3.

({\em b}) The renormalization prescriptions for Feynman amplitudes
(\ref{FPI}) are obtained form the above prescriptions for the products
$\Pi_{\Gm}(\ux,\uy)$ by 1) integrating over coordinates associated with
internal vertices
2) Fourier transforming in $y_l$ and putting the corresponding $m^2_l$
equal to squares of masses, e.g. $m_l=0$.
It is then natural to consider the insertion polynomials
${\cal P}_{\Gm}$
dependent only on derivatives in coordinates that correspond to
{\em external} vertices of the given graph.

\section{Commutativity of $R$-operation with \newline
differentiation and the action principle}

It is natural to check whether differential renormalization is in
agreement with the action principle which expresses basic properties of
quantum field theories such as equations of motion and the gauge
invariance.
Within dimensional renormalization the strategy for proving the
renormalized action principle \cite{BM} is
as follows: equations of motion, Ward identities etc.
are proved for unrenormalized
quantities, then  for regularized quantities
and finally (by more or less obvious commutativity of differentiation in
coordinates with renormalization) for renormalized quantities.
The most non-trivial point is to justify the relevant symmetries for
regularized quantities.

In the context of differential renormalization
there will be no such intermediate steps  following this program
because this is essentially
a renormalization without regularization.
To see what problems arise in reducing the problem to the case of
unrenormalized quantities (via commutation of differentiation with
respect to $R$-operation) let us consider the simplest example
of Fig.~1a:
\be R G(x)^2 = \hat{S} \ln \mu^2 x^2 G(x)^2 .\ee
Let us try to see what is
\[ \pr_{\al} R G(x)^2 ,\]
with $\pr_{\al} = \pr/\pr x_{\al}$.
First, remember that
both $\pr_{\al}$ and $\hat{S}$ are
understood in the distributional sense.
We have
\be \pr_{\al} \hat{S} = (\hat{S}+1/2) \pr_{\al} ,
\ee
and hence
\[ (\hat{S}+ 1/2) \pr_{\al}
\ln \mu^2 x^2 G(x)^2 . \]
Proceeding naively, for the moment, we continue to apply $\pr_{\al}$ using the
Leibniz rule. After this operation we obtain expressions that are
ultraviolet divergent.
However in a distributional sense the
derivative is not defined because  it acts on
{\em unrenormalized} expression that is defined only at $x\neq 0$.
This is a manifestation of an important difference with respect to
dimensional renormalization: we do not
have a regularization associated with differential
renormalization.

A different prescription is necessary in order to apply $\pr_{\al}$.
The point is that our
differential $R$-operation should take account of the degree of
divergence. For $G(x)^2$ it is equal to zero while formally it is
$\om=1$ for $\pr_{\al} G(x)^2$.
According to our prescriptions when the degree of
divergence $\om$ is equal to one
it is possible to use the operator (\ref{S1}) and
the corresponding differentially renormalized expression is
\be R \pr_{\al} G(x)^2
= -2 (\hat{S}+1/2) \hat{S} \ln \mu^2 x^2
\pr_{\al} G(x)^2.
\label{RDG}
\ee
Let us use the following equation:
\be
\hat{S} = - 2\hat{S} (\hat{S}-1/2) + 2 \hat{S}^2, \ee
with (\ref{F5}), (\ref{C1}) to show that
\be
(\hat{S}+1/2) \hat{S} \pr_{\al}
G(x)^2 = c_1 \pr_{\al} \dl(x). \ee
With the help of the auxiliary analytic regularization it is easy to
find
$-2c_1 = c_0 \equiv c_{1a}= 1/16\pi^2$.

We have
\[\pr_{\al} \hat{S} (\hat{S}-1/2) =
(\hat{S}+1/2) \hat{S} \pr_{\al}  \equiv\hat{S}^{(1)} \pr_{\al}.\]
Hence we may now apply $\pr_{\al}$ to $\ln \mu^2 x^2 /{x^4}$
naively, in the sense of ordinary functions rather than in a
distributional sense,
because the operator $\hat{S}^{(1)}$ ensures that the overall expression is
well defined.
As a result we see that commutativity breaks down:
\be \pr_{\al} \hat{S} \ln \mu^2 x^2 G(x)^2
= -2 (\hat{S}+1/2) \hat{S} \ln \mu^2 x^2
\pr_{\al} G(x)^2
+ \frac{3}{2} c_0 \pr_{\al} \dl(x).
\label{NC}
\ee
One can however define the renormalization of
$\pr_{\al} G(x)^2$ with another $\mu$-parameter, $\mu'$. This results in
\be \pr_{\al} \hat{S} \ln \mu^2 x^2 G(x)^2
= -2 (\hat{S}+1/2) \hat{S} \ln \mu'^2 x^2
\pr_{\al} G(x)^2
+ (3/2 -\ln(\mu'^2 / \mu^2) ) c_0 \pr_{\al} \dl(x).
\label{Co}
\ee
and so we may recover commutativity when $\ln(\mu'^2 / \mu^2) =3/2$.

This simple example shows that the commutativity of differentiation in
coordinates with the $R$-operation is not satisfied automatically in
differential renormalization.
It is necessary to adjust renormalization parameters to provide it.
Another possibility is to use desired commutation relations
as definitions for renormalization of diagrams that are obtained  as
derivatives of some other diagrams \cite{O}. In the above example,
this amounts to applying
$\pr_{\al} R G(x)^2$ (rather than the right-hand side of (\ref{RDG}))
as a definition of
$R \pr_{\al} G(x)^2$.

Bearing in mind this conclusion let us
consider, for example, the following equations in the $\phi^4$
theory:
\bea
m^2 \frac{\pa}{\pa m^2} RG^{(n)} = - R D_m G^{(n)}, \label{AP1} \\
g \frac{\pa}{\pa g} RG^{(n)} = - R D_4 G^{(n)}, \label{AP2} \\
R \left( D_m - D_2 + 2 D_4 + \frac{n}{2} \right) G^{(n)} = 0,
\label{AP3}
\eea
where $D_m G^{(n)}, D_2 G^{(n)}, D_4 G^{(n)}$
denote, respectively, $n$-point Green functions with insertions of
the following operators:
\be
\int \dd x \, m^2 \phi^2 (x)/2 , \; \int \dd x \, (\pa \phi)^2 (x) /2 , \;
\int \dd x \, g\phi^4 (x) /4!,
\label{VDO}
\ee
In dimensional renormalization these equations
directly follow from the
renormalized action principle \cite{BM} (in particular,
(\ref{AP3}) is the
equation of motion).

However
the renormalized action principle is not automatically guaranteed for
differential
renormalization. Nevertheless, it is possible to adjust finite
arbitrariness (at a diagrammatic level)
in renormalization of diagrams that contribute to the first two
operators in (\ref{VDO}) in such a
way that equations (\ref{AP1}--\ref{AP3}) hold.
In fact, it is sufficient to define renormalization of diagrams with one
insertion of operators $m^2\phi^2/2$
or $(\pa \phi)^2 /2$ to satisfy
\bea
R \int \dd x \, m^2 G(x-x_1) G(x-x_2) \overline{\Pi}(x_1,x_2, \ldots)
\nn \\
= -  m^2 \frac{\pa}{\pa m'^2}
R G(x_1-x_2;m')  \overline{\Pi}(x_1,x_2, \ldots) \big |_{m'=m} \; ,
\label{AP4} \\
R \int \dd x \left[ (m^2-\Box_x) G(x-x_1) \right] G(x-x_2)
\overline{\Pi}(x_1,x_2,\ldots)
\nn \\
= R G(x_1-x_2) \overline{\Pi}(x_1,x_2, \ldots),
\label{AP5}
\eea
where $\overline{\Pi}$ is the rest of the product of the propagators,
and in (\ref{AP4})
the mass derivative acts only on the first
propagator $G(x_1-x_2)$.
Since we certainly have possibility to adjust
coefficients of proportionality $\zeta_{\Gm}$ of the `overall'
$\mu$-parameter
$\mu_{\Gm} = \zeta_{\Gm} \mu$ to satisfy (\ref{AP4},\ref{AP5}), it is
sensible to use Eqs. (\ref{AP4},\ref{AP5}) as definitions of the
left-hand side.

\section{Renormalization group coefficients}

In dimensional renormalization the renormalization group
equation
\be
\left( \mu^2 \frac{\pa}{\pa \mu^2} +
\beta(g) \frac{\pa}{\pa g} - \gm_m(g) m^2 \frac{\pa}{\pa m^2}
+\frac{n}{2} \right) RG^{(n)} =0
\label{RGE}
\ee
can be derived \cite{C} from the so-called
diagrammatic RG equation
\be
-\mu^2 \frac{\pa}{\pa \mu^2} R F_{\Gm} + \eps h_{\Gm} R F_{\Gm} =
\sum_{\gm \subseteq \Gm}
h_{\gm} R \left( \Dl^{(1)} (\gm) F_{\Gm} \right) ,
\label{DRG}
\ee
where $h_{\gm}$ is the loop number,
the operation $\Dl^{(1)} (\gm) \equiv \eps
\hat{K}_{\eps}^{(1)} \Dl(\gm)$ is obtained from the
counterterm operation $\Dl^{\rm MS}(\gm)$ of the MS-scheme as the
residue of the simple pole in $\eps$.

Then one uses relations (\ref{AP1}--\ref{AP3}) and
obtains the well-known formulae for the RG coefficients:
\bea
\beta(g) & =  & g \frac{\pa Z_g^{(1)}}{\pa g} =
g(2\gm_2(g) - \gm_4(g)) ,
\label{RGC1} \\
\gm_{m^2}(g) & = & g \frac{\pa Z_{m^2}^{(1)}}{\pa g}
= g_2 (g) - \gm_{\phi^2}(g) ,
\label{RGC2} \\
\gm_{i}(g)&  = & - g \frac{\pa Z_{i}^{(1)}}{\pa g} , \; i=2,\phi^2,4 ,
\label{RGC3}
\eea
where $Z_i^{(1)}$ are contributions of simple poles to the
counterterms
\bea
Z_2 & = & 1+
\frac{\pa}{\pa p^2} \hat{K}_{\eps} R' G^{(2)}(p^2,m^2;\eps), \\
Z_{\phi^2} & = & 1+ \frac{\pa}{\pa m^2}
\hat{K}_{\eps} R' G^{(2)}(p^2,m^2;\eps), \\
Z_{4} & = & 1+\hat{K}_{\eps} R' G^{(4)}(p_1,\ldots,p_4,m^2;\eps).
\eea

In differential renormalization we have an equation similar to
(\ref{DRG}):
\be
-\mu^2 \frac{\pa}{\pa \mu^2} R F_{\Gm}
= \sum_{\gm \subseteq \Gm}
R \left( {\cal C}_\gm F_{\Gm} \right) ,
\label{DiRG}
\ee
where ${\cal C}_{\gm}$ is the operation (given by (\ref{CG}) and
(\ref{CGC}))
that inserts a finite polynomial of degree $\om$ in external momenta into the
reduced graph $\Gm / \gm$.
To prove (\ref{DiRG}) it is sufficient to repeat arguments
applied in proving (\ref{F18}). Moreover, (\ref{DiRG}) is in
agreement with homogeneity of renormalized Feynman amplitudes in
coordinates, inverse masses and $1/\mu$.

Under these conditions,
for evaluation of RG coefficients in differential renormalization,
we can apply the following formulae that are quite similar to
(\ref{RGC1}--\ref{RGC3}):
\bea
\beta(g)  & =  & c_g = g(2\gm_2(g) - \gm_4(g)) ,
\label{RGCD1} \\
\gm_{m^2}(g) &  = & c_{m^2}
= g_2 (g) - \gm_{\phi^2}(g) ,
\label{RGCD2} \\
\gm_{i}(g)  & =  & - c_i, \; i=2,\phi^2,4 ,
\label{RGCD3}
\eea
with
\bea
(c_{m^2} m^2-c_2\Box_x) \dl(x) =
{\cal C} RG^{(2)}(x), \label{FC1} \\
c_4 \prod_{i=2,3,4} \dl(x_i-x_1) = {\cal C} RG^{(4)}(x_1,\ldots,x_4).
\label{FC2}
\eea
The constants $c_i$ are calculated as sums of diagrammatic
contributions ${\cal C}_{\Gm} F_{\Gm}$.
Note that there is no factor $g \frac{\pa}{\pa g}$ in
(\ref{RGCD1}--\ref{RGCD3}) because, in contrast to (\ref{DRG}), Eq.
(\ref{DiRG}) does not involve the loop number $h_{\gm}$.

Thus, within differential renormalization, the constants
$c_i$ play the same role as the residues of the simple poles
in dimensional renormalization counterterms.

\section{Conclusion}

We have seen that the action principle in differential renormalization is
not satisfied automatically.
It is necessary to adjust renormalization parameters to satisfy
equations of motion etc. Moreover it is clearly natural and very
useful to employ equations of motion etc.
as definitions for renormalization of derivatives of the quantities
involved
(e.g. Green functions with insertions of composite operators) whenever
possible \cite{O}.

Based on subtractions on all divergent 1PI subgraphs differential
renormalization turns out to be a mass-independent scheme ---  this
property corresponds to locality of counterterms in the auxiliary
parameters $y_l$ as described in Sections 4 and 5. Renormalization group
calculations in this version of differential renormalization are rather
simple.
Since the problem reduces to
calculations of constants $c_{\gm}$ then one can use
the method of infrared rearrangement \cite{V} which is based on
possibility to put to zero masses and external momenta ($\equiv$
integration over some coordinates, from coordinate-space point of
view) --- see examples of such calculations in Sections 3 and 4.

An alternative approach to renormalization group calculations within
another version of differential renormalization \cite{O} is based, in
the massive case, on the short distance expansion of propagator in
coordinate space. This also results in a mass-independence property of
the renormalization. Up to two-loop order, such scheme is successfully
applied in various situations.
However, for higher orders, the relevant short-distance expansion of the
propagator should involve  many terms which can essentially complicate
the situation. It seems that the language of the auxiliary
$y$-variables is a necessary  price that must be paid
in order to have a practical
strictly four-dimensional scheme for multiloop calculations.
\vspace{3mm}

{\em Acknowledgements.}
I am very much grateful to K.G.~Chetyrkin and H.~Osborn
for valuable discussions.

\newpage
\setlength {\unitlength}{1mm}
\noindent {\Large \bf Figures}
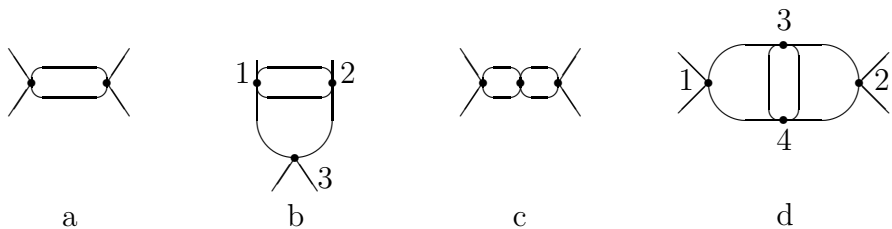
\begin{figure}[h]
\begin{picture}(150,28)
\put (10,20) {\circle*{1}}
\put (20,20) {\circle*{1}}
\put (15,20) {\oval(10,4)[b]}
\put (15,20) {\oval(10,4)[t]}
\put (20,20) {\line(2,3){3}}
\put (20,20) {\line(2,-3){3}}
\put (7,15.5) {\line(2,3){3}}
\put (7,24.5) {\line(2,-3){3}}
\put (14,1) {a}
\put (40,20) {\circle*{1}}
\put (50,20) {\circle*{1}}
\put (45,10) {\circle*{1}}
\put (45,20) {\oval(10,4)[b]}
\put (45,20) {\oval(10,4)[t]}
\put (45,20) {\oval(10,20)[bl]}
\put (45,20) {\oval(10,20)[br]}
\put (42,5.5) {\line(2,3){3}}
\put (45,10) {\line(2,-3){3}}
\put (40,20) {\line(0,1){3}}
\put (50,20) {\line(0,1){3}}
\put (37,20) {1}
\put (51,20) {2}
\put (48,6) {3}
\put (44,1) {b}
\put (70,20) {\circle*{1}}
\put (75,20) {\circle*{1}}
\put (80,20) {\circle*{1}}
\put (72.5,20) {\oval(5,4)[b]}
\put (77.5,20) {\oval(5,4)[b]}
\put (72.5,20) {\oval(5,4)[t]}
\put (77.5,20) {\oval(5,4)[t]}
\put (80,20) {\line(2,3){3}}
\put (80,20) {\line(2,-3){3}}
\put (67,15.5) {\line(2,3){3}}
\put (67,24.5) {\line(2,-3){3}}
\put (74,1) {c}
\put (100,20) {\circle*{1}}
\put (120,20) {\circle*{1}}
\put (110,15) {\circle*{1}}
\put (110,25) {\circle*{1}}
\put (110,20) {\oval(4,10)[l]}
\put (110,20) {\oval(4,10)[r]}
\put (110,20) {\oval(20,10)[b]}
\put (110,20) {\oval(20,10)[t]}
\put (96,16) {\line(1,1){4}}
\put (96,24) {\line(1,-1){4}}
\put (120,20) {\line(1,1){4}}
\put (120,20) {\line(1,-1){4}}
\put (96,19) {1}
\put (122,19) {2}
\put (109,27) {3}
\put (109,11) {4}
\put (109,1) {d}
\end{picture}
\caption{
Lower-order vertex diagrams from $\phi^4$-theory.}
\end{figure}
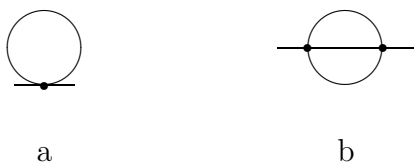
\begin{figure}[h]
\begin{picture}(150,28)
\put (40,20) {\circle{10}}
\put (40,15) {\circle*{1}}
\put (36,15) {\line(1,0){8}}
\put (39,5) {a}
\put (80,20) {\circle{10}}
\put (75,20) {\circle*{1}}
\put (85,20) {\circle*{1}}
\put (71,20) {\line(1,0){18}}
\put (79,5) {b}
\end{picture}
\caption{Lower-order self-energy diagrams from $\phi^4$-theory.}
\end{figure}
\end{document}